# Exploring the Effects of Data Set Choice on Measuring International Research Collaboration: an Example Using the ACM Digital Library and Microsoft Academic Graph

Ba Xuan Nguyen[1], Markus Luczak-Roesch[1] and Jesse David Dinneen[1]

[1] *ba.nguyen @vuw.ac.nz, markus.luczak-roesch@vuw.ac.nz, jesse.dinneen@vuw.ac.nz*
School of Information Management, Victoria University of Wellington, Wellington (New Zealand)

## Introduction

International research collaboration (IRC) is a construct that refers generally to scientific activities between individuals in different countries. IRC measurement is important because countries can and want to benefit from international collaboration (Guerrero Bote et al., 2013; Katz & Martin, 1997). Hence, ways to measure IRC are a focus of bibliometrics and informetrics research.

Many datasets are available to measure IRC and other facets of what has been framed as the "Science of Science" (Fortunato et al., 2018), but it has also been shown that performing the same measurement procedure on different datasets can lead to different results (De Stefano et al., 2013). The extent as well as the causes for such variances need to be adequately understood. We aim to contribute to this understanding by addressing the following research question: **what are the effects of data set choice on IRC measurement?**

## Research data and operationalisation of IRC

In this preliminary investigation we consider bibliographic metadata from the ACM Digital Library (ACM) and the Microsoft Academic Graph (MAG) dataset. The ACM data is supplied by the Association for Computing Machinery[1] as resource for research purposes (coverage from 1951-2017), while the MAG data (Sinha et al., 2015) was downloaded from the Open Academic Society website[2] (coverage from 1965-2017). Since ACM is largely a domain specific bibliographic source in the computing sciences (CS), a subset of the MAG dataset was created to cover only papers related to this field (by filtering with the most frequent "field of study" CS terms extracted from the matched collection). We acknowledge that applying this heuristic implies a limitation because it might mean we are missing some papers. In addition, some papers in this collection that already exist in ACM are also excluded to ensure that the ACM and MAG data sets are distinct.

In this study we investigate co-authored publication records and use the information about authors' affiliations to derive distinct **bilateral relationships**. If there is more than one co-author from a country in one publication, only one bilateral relationship is counted between that country and any of the others. From the ACM set of 182,791 records we identified 121,672 that are co-authored, 15,686 of which international co-authors, whereas from the MAG set of 939,821 computer science publications we found 594,036 with co-authors, of which 32,909 had international co-authors. This resulted in 21,827 (ACM) and 45,068 (MAG) distinct bilateral relationships.

## Analysis and results

First we observe a difference in the numbers of bilateral relationships between countries found in ACM and MAG. While the trend of both datasets in the last 10 years is similar, the MAG data shows a substantially different evolution compared to ACM before that time and has a significantly lower amplitude (see Figure 1).

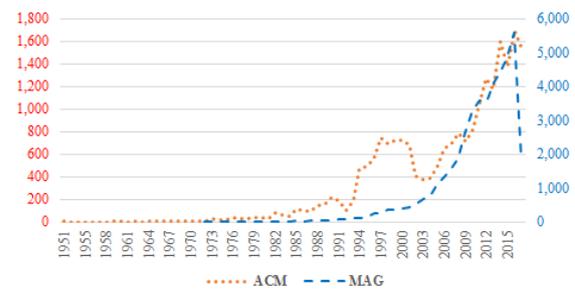

**Figure 1. Total numbers of bilateral relationships over 1951-2017.**

Comparing the top 10 countries ranked by total numbers of bilateral relationships over a period of 50 years (1966-2015), we find that the USA is consistently ranked first in both data sets. Other countries differ, however: for example China is ranked at the sixth position in the top ranks of ACM while being ranked second in MAG. Similarly, Canada is listed as the fourth in the former but sixth in the latter. To find out how consistent this ranking of countries is over time we perform an analysis of the rank order of countries based on international collaborations per year. To do this we first create a dataset of the annual IRCs per country (# of distinct countries: MAG - 164, ACM - 111), then derive an annual ranking of all countries by the amount of IRCs in the respective years (from highest to lowest), which we then inspect to find (a) the unique countries that are represented in both datasets (N=109) and (b) a reasonable cutoff point from which onwards we have a set of countries that are ranked in any of the following years. We set the cutoff to the year 2000, and derive a set of 30 countries that are fully covered from this year onwards until 2017, allowing us to rank these countries for all 18 years.

For each pair of rank vectors for these 34 countries we compute Spearman's and Kendall's rank order correlation coefficients, and the hamming distance. We also compute the mean and standard deviation (SD) for each of the rank vectors (so each country has one mean rank for its position in ACM and one for MAG). Again, the analysis of this data shows that the USA is consistently ranked first (and therefore has no correlation coefficients as the SD is zero). For all other countries we find that the mean of the hamming distance of the rank vectors is notably high (16.10, SD 2.37), which means that they are ranked differently in the two datasets for most years. Figure 2 displays the Spearman correlation coefficient (with confidence intervals) for these 30 countries. It highlights that there is basically no correlation present, which means that even the trend of the rank for countries (i.e. if a country rises over the years in one dataset it also rises in the other) is not consistent between the two datasets.

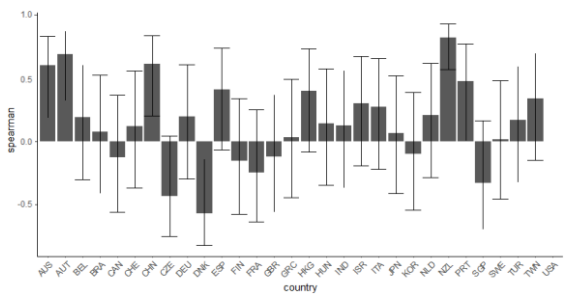

**Figure 2. Spearman's rank order correlation coefficients (with confidence interval) for 30 countries for rank vectors from 2000-2017.**

**Conclusion**

In this short paper we reported about our efforts to quantify and qualify the effects of data set choice on the outcomes of IRC measurement. We sought to provide empirical evidence that there are significant differences and to give some preliminary indicators for what cause and effect these may have. By performing an intuitive time series and rank order analysis **we found (1) inconsistent temporal coverage of the computer science domain in ACM and MAG data; (2) a similar upwards trend in bilateral scholarly relationships in recent years but with varying amplitude; and (3) significant movements in ranks for countries that are not consistent between the two datasets.**

We conclude that there exist differences that need to be better understood and require further investigation. The results presented here already have implications for our understanding of key activities in bibliographic analysis, such as temporal sampling when measuring IRC. Future work will need to dig deeper into the cause and effect relationships and we seek to undertake an analysis that clusters countries based on their rank variance to see if there are certain countries that are affected more or less than others by the differences in the data sets. Finally, the problem demonstrated here can also be looked at qualitatively to understand whether the data quality issues matter to actual data consumers such as policy makers.

---

[1] ftp://pubftp.acm.org

[2] https://www.openacademic.ai/oag/